\documentclass[aps,prb,twocolumn,psfig,showpacs]{revtex4}
\usepackage{amsfonts}
\usepackage{amsmath}
\usepackage{graphicx}
\usepackage{bm}
\usepackage{amssymb}
\usepackage{times}
\usepackage{dcolumn}
\usepackage{cases}
\usepackage{txfonts}

\begin{document}

\title{Phase Diagrams, Distinct Conformal Anomalies and
Thermodynamics of Spin-1 Bond-Alternating Heisenberg
Antiferromagnetic Chain in Magnetic Fields}
\author{Xin Yan$^\dagger$, Wei Li$^\dagger$, Yang Zhao, Shi-Ju Ran, and Gang Su}
\email[Corresponding author. ]{Email: gsu@gucas.ac.cn}
\affiliation{Theoretical Condensed Matter Physics and Computational
Materials Physics Laboratory, College of Physical Sciences, Graduate
University of Chinese Academy of Sciences, P. O. Box 4588, Beijing
100049, China}

\begin{abstract}

The ground state phase diagrams and thermodynamic properties of a
spin-1 bond-alternating Heisenberg antiferromagnetic chain with a
single-ion anisotropy in longitudinal and transverse magnetic fields
are investigated jointly by means of the infinite time evolving
block decimation, the linearized tensor renormalization group and
the density matrix renormalization group methods. It is found that
in the magnetic field-bond alternating ratio plane six phases such
as singlet dimer, Haldane, two Tomonaga-Luttinger liquid, 1/2
magnetization plateau and spin polarized phases are identified in a
longitudinal field, while in a transverse field there are fives
phases including singlet dimer, Haldane, Z$_2$ symmetry breaking,
quasi-1/2 magnetization plateau and quasi spin polarized phases. A
reentrant behavior of the staggered magnetization in a transverse
field is observed. The quantum critical behaviors in longitudinal
and transverse magnetic fields are disclosed to fall into different
universality classes with corresponding conformal field central
charge $c=1$ and $1/2$, respectively. The experimental data of the
compound NTENP under both longitudinal and transverse magnetic
fields are nicely fitted, and the Luttinger liquid behavior of low
temperature specific heat experimentally observed is also confirmed.

\end{abstract}

\pacs{75.10.Jm, 75.40.Mg, 05.30.-d, 02.70.-c}
\maketitle

\section{Introduction}

Low dimensional quantum spin systems have been active subjects in
quantum many-body physics for many years. Owing to strong quantum
fluctuations and competitions between various interactions in these
systems, a number of exotic and fascinating quantum emergent
phenomena are expected to occur, which thus arouses persistently
considerable interest not only in condense matter physics but also
in other fields such as quantum information and quantum computation \cite{andrew}.

Among others, the spin-1 bond alternating Heisenberg
antiferromagnetic chain (BAHAFC) in longitudinal and transverse
magnetic fields is of particular interest, for a series of
materials, such as NENP \cite{S.Ma}, NMOAP \cite{Narumi}, NDOAP
\cite{Narumi2}, and NTENP \cite{escuer} can be well described by the
spin S=1 BAHAFC, and the ideal model compound
Ni(C$_{9}$H$_{24}$N$_{4}$)(NO$_{2}$)ClO$_{4}$ (NTENP) has been
extensively studied both experimentally and theoretically in recent
years
[\onlinecite{Narumi,Narumi3,Narumi2,Zheludev,Hagiwara,Suzuki,regnault,Hagiwara2,glazkov}],
where the effects of magnetic field and the bond alternating ratio
on low-lying magnetic excitations, spin correlations and
low-temperature specific heat of the compound NTENP have been
investigated. It is known that in a magnetic field the
one-dimensional (1D) Heisenberg quantum spin chains with periodic
ground states would exhibit a topological quantization of
magnetization according to the necessary condition $n(S-m)=integer$
\cite{Oshikawa}, where $n$ is the period of the system, $S$ is the
spin and $m$ is the magnetization per site. Such magnetic plateau
states have been addressed in some polymerized Heisenberg
antiferromagnetic or ferrimagnetic spin chains (e.g. Refs.
[\onlinecite{su}]). High-field magnetization measurements on NTENP
also revealed that a $m=1/2$ magnetization plateau appears around
700 kOe \cite{Narumi,Narumi2}; the neutron scattering experiments
indicated that the spin dynamics of NTENP quite differs from that of
a Haldane spin chain \cite{Zheludev,Hagiwara,glazkov}; and the
low-temperature specific heat measurement showed a
Tomonaga-Luttinger liquid (TLL) behavior in NTENP in a longitudinal
magnetic field \cite{Hagiwara2}. Although the experimental studies
on NTENP gain obvious advances, some ambiguities still remain on
further understanding its physical properties. For instance, the
compound NTENP was experimentally studied under both longitudinal
and transverse magnetic fields, but only the data in a longitudinal
field have been fitted by quantum Monte Carlo simulations; no direct
numerical evidence of low-temperature specific heat of this material
was reported to verify the TLL behavior; the sharp peaks of the
specific heat in the NTENP were observed experimentally in both
longitudinal and transverse magnetic fields, where the peak
positions move to high temperature side with the increase of the
transverse field, while retain almost intact with increasing the
longitudinal field; and so on. The remaining issues are very worthy
to address.

Apart from that the spin-1 BAHAFC with a single-ion anisotropy [Eq.
(\ref{eq-hamiltonian}) below] is believed to be a pertinent model in
describing the physical characters of NTENP, which has been studied
previously \cite{Zheludev,Gomez,Tsvelik,Kitazawa, Singh, Kohno},
this model itself has also fascinating properties that deserve to
investigate. When the bond alternating ratio $\alpha=0$, it reduces
to the singlet dimers; when $\alpha=1$, it becomes a spin-1 uniform
chain (Haldane chain). In these two special cases, the system shows
an excitation gap from the singlet dimer or Haldane ground state to
the triplet excited state. The latter uniform chain under both
longitudinal ($h_z$) and transverse ($h_x$) magnetic fields has been
extensively discussed (e.g. Refs.
[\onlinecite{Affleck,X.G.Wen,Xing}]). However, the overall phase
diagrams in $\alpha-h_{z,x}$ plane for the S=1 BAHAFC with a
single-ion anisotropy are still absent. Therefore, it is really
necessary to tackle these above questions.

In this paper, by means of jointly the infinite time evolving block
decimation (iTEBD) \cite{G. Vidal}, the linearized tensor
renormalization group (LTRG) \cite{Li}, and the density matric
renormalization group (DMRG) methods \cite{White}, we shall study
systematically the ground state phase diagrams, magnetic and
thermodynamic properties of the S=1 BAHAFC with a single-ion
anisotropy in longitudinal and transverse magnetic fields. The
iTEBD, LTRG and DMRG methods are the numerical algorithms with very
good accuracy and high efficiency that were recently developed for
low-dimensional quantum lattice systems, which allow for calculating
with nice precision the critical properties and the extremely low
temperature behaviors of quantum many-body lattice systems. These
methods can assist us to fit well the experimental data on the NTENP
with the accurately calculated results, thereby capable of
reasonably determining the material parameters and better
understanding the fundamental features of the NTENP in both
longitudinal and transverse magnetic fields.

\section{Model and Ground State Phase Diagrams}

\subsection{Model Hamiltonian and Calculational Method}

Let us start with the model Hamiltonian given by
\begin{eqnarray}
H &=& J \sum_i^{L/2} (\bold{S}_{2i-1} \cdot \bold{S}_{2i} + \alpha
\bold{S}_{2i} \cdot \bold{S}_{2i+1}) \notag \\
&+& \sum_i^L [ \Delta (S_i^z)^2 -g_\| \mu_B h_z S_i^z -g_\bot \mu_B
h_x S_i^x], \label{eq-hamiltonian}
\end{eqnarray}
where $J$ is the coupling constant, $L$ (even) is the number of
lattice sites, $\bold{S}_{i}$ is the spin operator at $i$th site,
$\Delta$ is the single-ion anisotropy, $g_\|$, $g_\bot$ are the
Land\'e g-factors, $\mu_B$ is the Bohr magneton, $h_z$ and $h_x$ are
the longitudinal and transverse magnetic fields, respectively.

In the following calculations, $g_{\|}\mu_B=g_{\bot}\mu_B=1$ is
assumed unless the fittings to experimental data are concerned, and
$J$ is taken as the energy scale. The ground state properties were
studied by utilizing the iTEBD \cite{G. Vidal} imaginary time
projection scheme along with DMRG method \cite{White}, and the
thermodynamic properties and fittings to the experimental data of
NTENP were performed by invoking the LTRG approach \cite{Li}. In
iTEBD calculations we take the smallest Trotter step $\tau=10^{-7}$.

\subsection{Magnetization and Phase Diagram in a Longitudinal Field}

\begin{figure}[tbp]
\includegraphics[width=1.0\linewidth,clip]{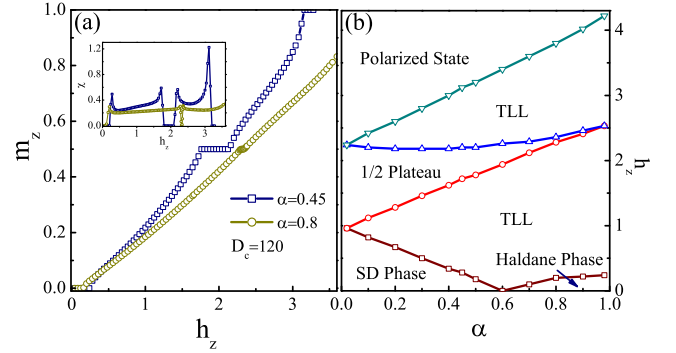}
\caption{(Color online) (a) The magnetic curves of spin-1 bond
alternating Heisenberg antiferromagnetic chain (BAHAFC) with a
single-ion anisotropy in longitudinal magnetic fields $h_z$, where
three magnetization plateaux ($m=0$, 1/2, and 1) are observed. The
inset shows the singularities of the susceptibility as a function of
magnetic field, which signal the occurrence of quantum phase
transitions. (b) The ground state phase diagram in the plane of the
bond alternating ratio ($\alpha$) \emph{vs.} longitudinal magnetic
field ($h_z$), where the Haldane, singlet-dimer (SD), m=1/2 magnetic
plateau, Tomonaga-Luttinger liquid (TLL), and spin polarized phases
are identified. The single-ion anisotropy $\Delta=0.25$ is taken.}
\label{fig-long-field-diag}
\end{figure}

Figure 1(a) presents the field dependence of magnetization per site
$m_z$ for the system defined by Eq. (\ref{eq-hamiltonian}) in the
longitudinal magnetic field for $\alpha=0.45$ and $0.8$, where the
bond dimension $D_c=120$ of a matrix product state is presumed. When
$\alpha=0.45$, it is clear to see that there are three magnetization
plateaux occurring at $m$=0, 1/2 and 1 in some regions of $h_z$. The
$m$=0 plateau suggests that a gap remains in the absence of magnetic
field, implying that the system is in the singlet-dimer phase, as it
should be in the same phase as $\alpha=0$. Based on the valence-bond
solid picture, the $m=1/2$ plateau corresponds to that one of the
two bonds in each coupled spin-1 dimer is broken \cite{su}. The
$m=1$ plateau is in the spin fully polarized state. Near the
critical fields, the magnetization per site $m_z$ depends on the
magnetic field in a square root behavior, similar to the case of
spin-1 HAFC in a magnetic field. In the ranges between the plateaus,
the spin correlations should be in a power law decay, indicating a
quasi long-range order. The critical magnetic fields for closure of
the gap can be accurately determined by the field dependence of the
susceptibility, where the sharp peaks appear, as shown in the inset
of Fig. 1(a). When $\alpha=0.8$, the three magnetic plateaux still
remain, but the width of $1/2$ plateau decreases, indicating that
the corresponding gap becomes narrower. For the spin-1 BAHAFC, as
the ground state period is $n=2$, the topological quantization
condition $n(S-m)=integer$ is satisfied by $m$=0, 1/2 and 1, i.e.,
there should exist at most three plateaux in the longitudinal field
for $0\leq \alpha <1$, which is well confirmed in Fig. 1(a). At the
special case of $\alpha=1$, $n$ becomes 1, and there should be two
plateaus at $m$=0 and 1.

For $0\leq \alpha \leq 1$, we have calculated the magnetization
curves and the corresponding susceptibilities in the ground state,
and collected all critical values of magnetic fields at which the
susceptibility exhibits obvious singularities, which allows us for
readily obtaining a global ground state phase diagram in
$\alpha-h_z$ plane for the spin-1 BAHAFC with a single-ion
anisotropy $\Delta=0.25$, as shown in Fig. 1(b). There are six
phases, including the singlet-dimer (SD) phase, Haldane phase, two
TLL phases separated by the $m=1/2$ magnetic plateau phase, and the
spin fully polarized phase. At a critical bond alternating ratio
$\alpha_c\approx0.6$, there exists a quantum phase transition from
the gapped SD state into the gapped Haldane phase. The $m$=1/2
plateau state persists for all values of $\alpha$ until it
disappears at $\alpha=1$. At all six phase boundaries, the gap
closes. In these phases, the spin-spin correlation function $\langle
S_i^zS_j^z\rangle$ reveals different spatial behaviors, which is
short-range ordered except that in the TLL phase it decays in an
algebraic way.

\subsection{Magnetization and Phase Diagram in a Transverse Field}

\begin{figure}[tbp]
\includegraphics[width=1.0\linewidth,clip]{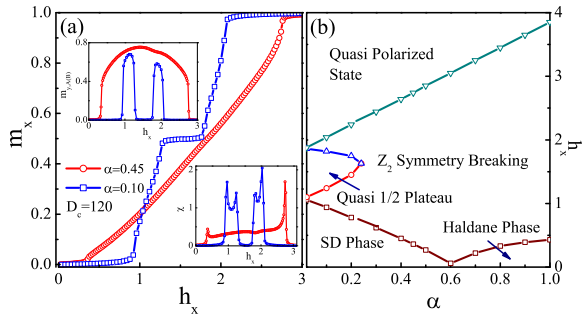}
\caption{(Color online) (a) The magnetic curves of S=1 BAHAFC with a
single-ion anisotropy in transverse magnetic fields $h_x$. The
up-left inset shows the staggered magnetization along the $y$ axis,
and the down-right inset presents the corresponding susceptibility
as a function of magnetic field. For $\alpha=0.1$, three
plateau-like steps are observed. (b) The ground state phase diagram
in the plane of the bond alternating ratio ($\alpha$) \emph{vs.}
transverse magnetic field ($h_x$), where the Haldane, singlet-dimer
(SD), Z$_{2}$ symmetry breaking N\'{e}el ordered, quasi 1/2 magnetic
plateau, and quasi spin polarized phases are identified. The
single-ion anisotropy $\Delta=0.25$ is taken.}
\label{fig-trans-field-diag}
\end{figure}

In a transverse magnetic field $h_x$, the field dependence of
magnetization per site $m_x$ and $m_y$ and the corresponding
susceptibility of the model under interest are presented in Fig.
2(a) for $\alpha=0.1$ and 0.45. Three steps around $m_x$=0, 1/2 and
1 in magnetic curves ($m_x$ \emph{vs.} $h_x$) are observed for
$\alpha=0.1$. The corresponding susceptibility $\chi$ is illustrated
in the lower inset of Fig. 2(a), where one may notice that, due to
the non-conservation of total $S_x$, at these three steps $\chi$
does not exactly vanish although it is negligibly small. Thus, we
call these steps as quasi magnetic plateaus. In the regions between
the quasi plateaus, $m_y$ has large values as shown in the upper
inset of Fig. 1(a). The reason is that the existence of the
single-ion anisotropy makes the spins tend to arrange in $xy$ plane,
and when the magnetic field is applied along the $x$ direction, the
staggered magnetization $m_y$ is induced to form a canted Ising
order. In addition, for $\alpha=0.1$ we observed from the upper
inset of Fig. 2(a) that with increasing $h_x$ the staggered
magnetization $m_y$ first is negligible small, and then exhibits a
sharp valley structure, which illustrates a reentrant behavior of
the staggered magnetization \cite{Diep,Hieida}. For $\alpha=0.45$,
only two steps appear in the magnetic curve, and the quasi $m=1/2$
step disappears. In this case, $m_y$ has even larger values in the
region between the two steps. For other larger $\alpha$, the
magnetic curves in the transverse field $h_x$ have the behaviors
similar to that of $\alpha=0.45$.

Utilizing the method similar to Fig. 1(b) and sweeping various
values of $\alpha$, we can collect all critical magnetic fields by
finding the singular positions of susceptibility, and then obtain
the whole ground state phase diagram in $\alpha - h_x$ plane for the
system under investigation, as depicted in Fig. 2(b). It can be seen
that there are five phases, namely, the SD, Haldane, quasi $m=1/2$
magnetic plateau, Z$_2$ symmetry breaking (or spin canted) and the
quasi spin polarized phases. The quantum critical point
$\alpha_c\approx0.6$ is recovered here. In contrast to the case
under a longitudinal field where the $m=1/2$ plateau phase persists
into the whole region of $\alpha<1$ and separates two TLL phases,
the quasi $m=1/2$ plateau phase sets only in a small region ($\alpha
\leq 0.24$). Starting from the SD and Haldane phases, when we
increase the field $h_x$, the gap is gradually closed at the phase
boundaries. If one continues to increase the magnetic field $h_x$,
another gap opens because of the breaking of the discrete symmetry,
giving rise to a Z$_2$ symmetry breaking phase, which displays a
long range behavior.

\section{Entanglement Entropy and Conformal Anomalies}

\begin{figure}[tbp]
\includegraphics[angle=0,width=1.0\linewidth,clip]{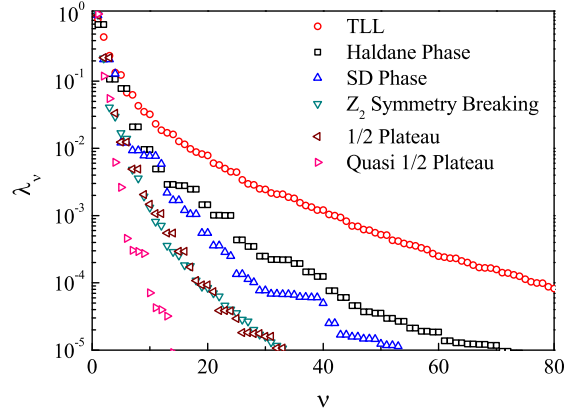}
\caption{(Color online) The normalized Schmidt coefficient
$\lambda_{\nu}$ as a function of the number of kept states $\nu$ for
different phases, where $D_{c}=80$. } \label{fig-spectrum}
\end{figure}

This present system demonstrates quite different behaviors in
longitudinal and transverse magnetic fields. To gain further insight
into the underlying physics behind this character, we have studied
the conformal anomalies of this system at the critical regimes. The
conformal field theory (CFT) tells us that the conformal invariance
at the critical point sets useful constraints on the critical
behaviors of two-dimensional classical or 1D quantum systems
\cite{conformal}, and the universality class can be characterized by
the conformal anomaly or central charge $c$ of the Virasoro algebra.
For a system with a continuous symmetry $G$, if G=SU(N), the
possible conformal central charge is given by $c=(N^2-1)k/(N+k)$,
$k=1,2,3,...$ \cite{Affleck0}. For the spin-S uniform
antiferromagnetic quantum chains, G=SU(2), $c=3S/(1+S)$. For S=1/2,
$c=1$; S=1, $c=3/2$. For the present spin-1 BAHAFC with a single-ion
anisotropy, the SU(2) symmetry is not satisfied, and the conformal
central charge should be calculated via other ways. The von Neumann
entropy $S$, defined by
\begin{equation}
S = -Tr(\hat{\rho}_{sys} \log_2 \hat{\rho}_{sys}) =
-Tr(\hat{\rho}_{env} \log_2 \hat{\rho}_{env}),
\label{entangle-entropy}
\end{equation}
offers a possible way to get the central charge of quantum spin
chains, where $\hat{\rho}_{sys (env)}$ is the reduced density matrix
(DM) of system (environment). In critical and noncritical regimes,
the entanglement entropy has different asymptotic behaviors
\cite{arealaw,Vidal2}. In critical regimes, the CFT predicts
\cite{wilczek}
\begin{equation}
S \approx \frac{c}{3} \log_2 (L) + k, \label{asymptotic}
\end{equation}
where $L$ is the number of spins for a block embedding in an
infinite chain, $c$ is the central charge and $k$ is a non-universal
constant. In noncritical regimes, $S$ vanishes for all L or grows
monotonically with L until saturation. In the following, we shall
invoke the iTEBD method to study the entanglement entropy for the
case in a transverse magnetic field. Because of the nearly unitary
evolution, the iTEBD algorithm gives the canonical form of infinite
matrix product states (iMPS). We can make a Schmidt decomposition on
the ground state wave function such that
\begin{equation}
 |\psi\rangle = \sum_{\nu=1}^{D_c}\lambda_{\nu}|\psi^{A}_{\nu}\rangle|\psi^{B}_{\nu}\rangle
\end{equation}
on each bond, where $|\psi_{\nu}^{A,B}\rangle$ is the
orthonormalized basis states (Schmidt bases) for two parts $A$ and
$B$ of the infinite chain, and $\lambda_{\nu}$ is the Schmidt
coefficient (SC). Fig. \ref{fig-spectrum} presents $\lambda_{\nu}$
as a function of $\nu$ for different phases. It is seen that in the
TLL phase, the SC attenuates much slower than those in
\begin{figure}[tbp]
\includegraphics[angle=0,width=1.0\linewidth,clip]{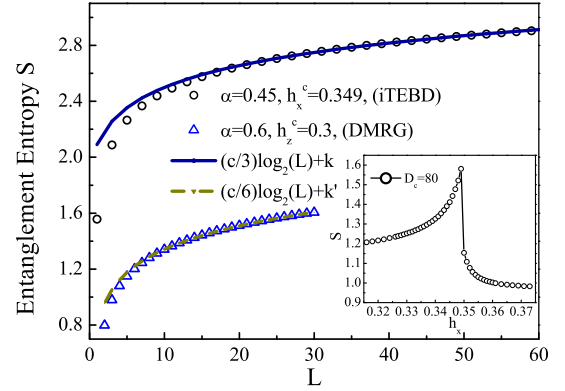}
\caption{(Color online) The entanglement entropy $S$ as a function
of chain length $L$ calculated by the iTEBD in a transverse field
for $\alpha=0.45$ and $D_{c}=100$, and by the DMRG in a longitudinal
magnetic field for $\alpha=0.6$ where the optimal states were kept
as 400. In both cases, the single-ion anisotropy $\Delta=0.25$. The
solid line is the fitting curve to Eq. (3), giving the central
charge $c = 1/2$ at a critical transverse field $h_{x}^c=0.349$; the
inset shows that the entanglement entropy as a function of $h_x$ for
a semi-infinite chain length is singular at $h_{x}^c=0.349$. The
dashed line is the fitting curve for the case (TLL phase) at a
longitudinal magnetic field $h_z=0.3$, giving the central charge
$c=1$.} \label{fig-central-charge}
\end{figure}
other gapped phases. The double degeneracy of SC only appears in the
Haldane phase, indicating the existence of the topological string
order \cite{entang_spectrum}. Given a canonical form of iMPS, if the
system in Eq. (\ref{entangle-entropy}) is chosen as the
semi-infinite chain, it is easy to directly calculate the von
Neumann entropy, $S = -
\sum_{\nu=1}^{D_{c}}\lambda_{\nu}^{2}\log_{2}(\lambda_{\nu}^{2})$.
In addition, we can also calculate the entanglement entropy when the
system is successive L spins embedding in an infinite chain, where
we may set $L = 2$ for an example. In particular, we can write
\begin{equation}
 \displaystyle|\psi\rangle = \sum_{\alpha,\beta,\gamma = 1}^{D_{c}}\sum_{i,j = 1}^{d}\lambda_{\alpha}\Gamma^{i}_{\alpha,\beta}\lambda_{\beta}\Gamma^{j}_{\beta,\gamma}\lambda_{\gamma}
 |\Phi_{\alpha}\rangle|i\rangle|j\rangle|\Phi_{\gamma}\rangle,
\end{equation}
where $\Gamma^{i}_{\alpha,\beta}$ is the tensor on the $i$-th site,
$|i\rangle (|j\rangle)$ is the spin basis states on the $i(j)$-th site,
$|\Phi_{\alpha}\rangle$ and $|\Phi_{\gamma}\rangle$ are the left
Schmidt basis of the $i$-th site and the right Schmidt basis of the
$j$-th site, respectively. By contracting two tensors, we will
obtain the reduced DM
\begin{eqnarray}
 \hat{\rho} &=& Tr_{ij}|\psi\rangle\langle\psi| \nonumber \\
 &=&\sum_{\alpha,\gamma=1}^{D_{c}}\sum_{\alpha',\gamma'=1}^{D_{c}}[\sum_{i,j=1}^{d}\sum_{\beta,\beta'=1}^{D_{c}}
 \lambda_{\alpha}\Gamma^{i}_{\alpha,\beta}\lambda_{\beta}\Gamma^{j}_{\beta,\gamma}\lambda_{\gamma}\nonumber \\
 & &\lambda_{\alpha'}(\Gamma^{i}_{\alpha',\beta'})^*\lambda_{\beta'}(\Gamma^{j}_{\beta',\gamma'})^*\lambda_{\gamma'}]
 |\alpha\rangle|\gamma\rangle\langle\alpha'|\langle\gamma'|.
\end{eqnarray}
One can diagonalize this matrix and calculate the entanglement
entropy. For a larger $L$, we need to contract $L$ tensors and
select Schmidt basis to construct the reduced DM. For all $L$, the
dimension of the reduced DM is $D_{c}^{2}\times{D_{c}^{2}}$. By
calculating the von Neumann entropy for a semi-infinite chain in a
transverse magnetic field, we observe that the cusp position of the
entanglement entropy just gives the phase transition point
$h_{x}^{c}=0.349(2)$. By fitting our calculated results to Eq.
(\ref{asymptotic}), we find $c = 1/2$ for $h_{x}^{c}=0.349$, as
shown in Fig. \ref{fig-central-charge}. In the TLL phase of the case
in a longitudinal magnetic field, the iTEBD algorithm gives the
ground state with a staggered magnetization perpendicular to $z$
direction so that it cannot yield a correct wave function. The
reason is that there does not have an excitation gap and the iTEBD
algorithm cannot project the right ground state wave function. In
order to calculate the central charge in the TLL phase, we utilize
the DMRG algorithm with open boundary condition at $h_{z}=0.3$,
$\alpha=0.6$, and $\Delta=0.25$. The result is also included in Fig.
\ref{fig-central-charge}, where the fitting result gives the central
charge $c=1$ in this TLL phase \cite{note}. Consequently, this
implies that the critical behaviors of the spin-1 BAHAFC with a
single-ion anisotropy in transverse fields quite differ from those
in longitudinal fields, and the universality falls into two distinct
classes with conformal central charges $c=1/2$ and $c=1$,
respectively.

\section{Thermodynamic Properties and Comparison to Experiments}

\subsection{Susceptibility, Magnetization and Comparison to Experiments}

\begin{figure}[tbp]
\includegraphics[angle=0,width=1.0\linewidth,clip]{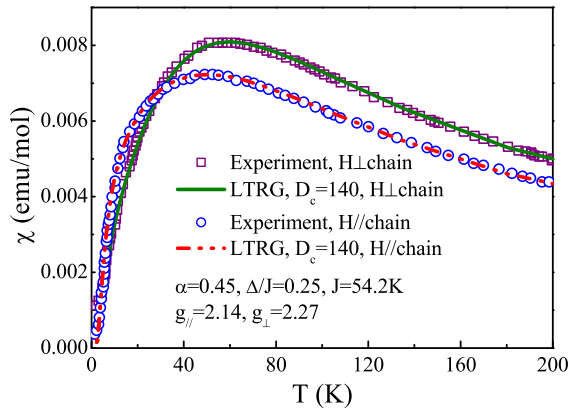}
\caption{(Color online) The temperature dependence of susceptibility
of NTENP measured experimentally is well fitted to the LTRG
calculated data for both longitudinal and transverse magnetic
fields. The experimental data are taken from Ref.
[\onlinecite{Narumi2}].} \label{fig-mag-sus}
\end{figure}

The thermodynamic properties of this system were explored by using
recently proposed LTRG algorithm \cite{Li} that allows for
accurately calculating the thermodynamic quantities at very low
temperature. In the following LTRG calculations we keep the Trotter
step $\tau=0.1$.

\begin{figure}[tbp]
\includegraphics[angle=0,width=1.0\linewidth,clip]{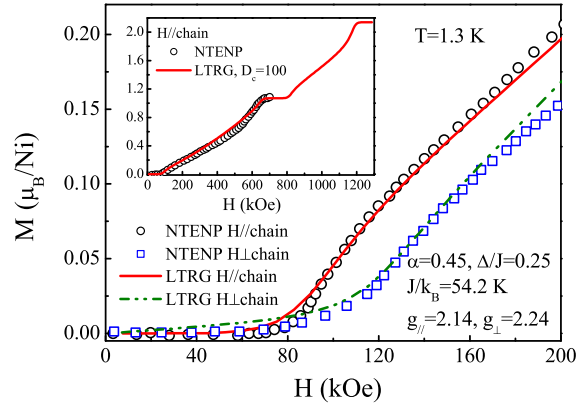}
\caption{(Color online) The magnetic curves of NTENP are well fitted
to the LTRG calculated data for both longitudinal and transverse
fields, where the fitting parameters are consistent with those from
the susceptibilities. Inset is the high-field magnetic curve up to
700 kOe fitted with the corresponding LTRG calculated data. The
experimental data are taken from Ref. [\onlinecite{Narumi2}].}
\label{fig-mag-curv}
\end{figure}

Fig. \ref{fig-mag-sus} gives the temperature dependence of
susceptibilities of the spin-1 BAHAFC in longitudinal and transverse
magnetic fields, where the fittings to the experimental data of
NTENP under both fields are also included. One may see that the
theoretical results are nicely fitted to the experimental data,
generating a set of material parameters of NTENP: $\alpha=0.45$,
$\Delta/J=0.25$, $J=54.2 K$, $g_{\|}=2.14$, and $g_\bot =2.27$. It
should be noted that in a longitudinal field, these parameters are
in agreement with the previous results from the quantum Monte Carlo
calculations \cite{Narumi2}, while in a transverse field the fitting
to the experimental data of NTENP is for the first time done. The
magnetization curves of NTENP were measured at T=1.3 K up to 700 kOe
in Ref. [\onlinecite{Narumi2}], which are well fitted to our LTRG
calculated results, giving the same set of fitting parameters as
those obtained from the data of susceptibility except that
$g_\bot=2.24$ here, as shown in Fig. \ref{fig-mag-curv}. Considering
that the two sets of experimental data from susceptibility and
magnetic curves are independent, a slight difference on $g_\bot$ is
reasonable. Thus, $g_\bot$ of NTENP should be around 2.24 $\sim$
2.27. The high-field magnetization curve is also well fitted with
our LTRG results (inset of Fig. (\ref{fig-mag-curv})) using the same
parameters.

\subsection{Specific Heat in Magnetic Fields}

\begin{figure}[tbp]
\includegraphics[angle=0,width=1.0\linewidth,clip]{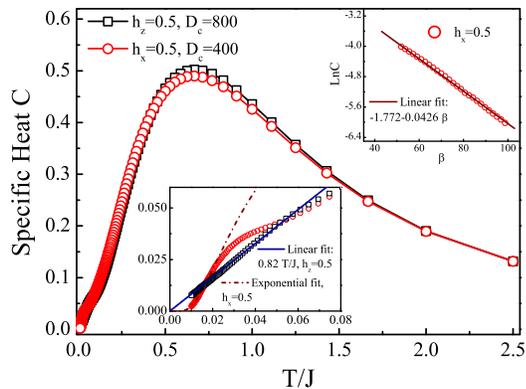}
\caption{(Color online) The temperature dependence of the specific
heat of S=1 BAHAFC under both longitudinal and transverse magnetic
fields. Although the whole profiles look similar for both cases, the
extremely low temperature behaviors shown in the down-left inset
quite differ, where the linear-T dependence of the specific heat in
a longitudinal field is clearly seen at low temperature, suggesting
a Tomonaga-Luttinger liquid behavior, while an exponential decay for
the T-dependence of the specific heat is observed in a transverse
field owing to the appearance of an Ising gap, as seen in the
up-right inset. The single-ion anisotropy $\Delta=0.25$ and the bond
alternating ratio $\alpha=0.45$ are taken.} \label{fig-spec-heat}
\end{figure}

The temperature dependence of the specific heat $C(T)$ of the S=1
BAHAFC in longitudinal and transverse magnetic fields is obtained by
the LTRG method down to very low temperatures, as shown in Fig.
\ref{fig-spec-heat} for $h_z=0.5$ and $h_x=0.5$. In a transverse
field, there are one round peak and a low temperature shoulder in
$C(T)$, while in a longitudinal field the specific heat exhibits
only one broad peak. It appears that at low temperature the specific
heat displays distinct behaviors. To show this point clearly, we
have carefully calculated the specific heat of this model at
extremely low temperatures by the LTRG algorithm that is very
powerful and efficient for calculating the low-temperature
properties than other numerical methods \cite{gubo}. Shown in the
lower inset of Fig. \ref{fig-spec-heat}, two different behaviors in
both fields at low temperature are clearly demonstrated, where in a
longitudinal field, the specific heat displays a linear T-dependent
character, showing a TLL behavior, while in a transverse field,
$C(T)$ exhibits an exponential decay (the fitting curve is given in
the upper inset of Fig. \ref{fig-spec-heat}), that can be ascribed
to the $Z_2$ symmetry breaking with an open of an Ising gap.

The specific heat $C(T)$ of NTENP was also measured experimentally
\cite{Hagiwara2}, where the linear T-dependence of $C(T)$ at low
temperature in a longitudinal field above the critical field
$h{_z}^c=9.3$ T was observed, which was identified as a TLL
behavior. Our low temperature LTRG calculations at $h_z=0.5$ ($ >
h{_z}^c$) on the S=1 BAHAFC model strongly supports this
experimental observation. In a transverse field, the experiment on
NTENP gives a distinct nonlinear low-temperature behavior of $C(T)$
from that in a longitudinal field, which is also backed up by our
LTRG results. We should remark here that the sharp peaks in the
temperature dependent specific heat of NTENP were observed in both
fields, which cannot be explained by using this spin-1 BAHAFC model,
as revealed by our present studies, because we do not find any
field-dependent sharp peaks of $C(T)$ in this 1D model. Those sharp
peaks may signal the field-induced long-range orders from the 3D
effect of inter-chain interactions in NTENP. It is this reason that
makes us not directly fit our LTRG results with the experimental
data of specific heat of NTENP. Nevertheless, our present studies on
the spin-1 BAHAFC model may give a possible clue to understand the
experimentally observed low-temperature sharp peaks of $C(T)$
\cite{Hagiwara2}. In a longitudinal field, as long as the magnetic
field is higher than the critical field ($h{_z}^c$), the system will
enter the TLL phase that has a quasi LRO, and the true LRO can be
established only by interchain couplings. Thus the peak position is
almost determined by the strength of interchain interactions and
hardly moves with the increase of the magnetic field. In a
transverse field, in the range that the experiment was performed,
the staggered magnetization that could enhance the interchain
couplings increases monotonously with increasing the magnetic field,
so the low-temperature sharp peak moves to the high temperature side
with the increase of the transverse field.

\section{Summary and Conclusion}

In summary, we have investigated the spin-1 BAHAFC model with a
single-ion anisotropy in longitudinal and transverse magnetic fields
by employing the iTEBD and LTRG methods. The ground state phase
diagrams in the plane of field versus the bond alternating ratio
under both fields are obtained, where various phases are identified
for two cases. From the entanglement entropy, the conformal central
charges in both critical longitudinal and transverse fields are
determined to be $c=1$ and $1/2$, respectively, suggesting that the
universality in critical regimes falls into different classes for
both fields. The TLL behavior at low $T$ observed experimentally in
NTENP is verified via our accurate calculations, while an
exponential decay of low-temperature specific heat is uncovered in a
transverse field. The experimental data of the model material NTENP
are well fitted with our LTRG results, and the parameters for
characterizing NTENP are determined.

\acknowledgments

We are indebted to Bo Gu, Fei Ye, Shou-Shu Gong, Bin Xi, J. Sirker,
and Qing-Rong Zheng for stimulating discussions. This work is
supported in part by the NSFC (Grants No. 10934008, No. 90922033),
the MOST (Grant No. 2012CB932900) and the Chinese Academy of
Sciences.

\end{document}